\documentclass[
PRR,
superscriptaddress,
longbibliography,
twocolumn,
showpacs,
amsmath,
amssymb,
aps,
floatfix
]{revtex4-2}

\usepackage{blindtext}
\usepackage{graphicx}
\usepackage{svg}
\usepackage{gensymb}
\usepackage{amsmath}
\usepackage{lipsum}  

\begin{document}
\title{Anisotropic magnetism at the surface of a non-magnetic bulk insulator}
\author{Jarryd A. Horn}
\author{Keenan E. Avers}
\author{Nicholas Crombie}
\author{Shanta R. Saha}
\author{Johnpierre Paglione}
\affiliation{Maryland Quantum Materials Center and Department of Physics, University of Maryland, College Park, Maryland 20742, USA}

\begin{abstract}
The potential for topological Kondo insulating behavior in $d-$electron systems has attracted interest in studying the surface states of the correlated insulators FeSb$_2$ and FeSi. While detailed studies and theoretical description of a spin-orbit coupled ferromagnetic surface state have been applied to FeSi, the magnetic properties of the surface states of FeSb$_2$ have not been addressed. Here, we report on the surface magnetic properties of FeSb$_2$, utilizing the surface area dependence of magnetic susceptibility to separate the surface Curie-Weiss temperature dependence from the bulk spin-gap susceptibility. We use these results to further extract the surface magnetic anisotropy of a thin, rough-surfaced single-crystal FeSb$_2$ to compare with the observed magnetotransport anisotropy, and find good agreement between the anisotropy in the surface magnetization and surface magnetotransport. {\color{black} Finally, we report a nonlinear component of the low-temperature surface Hall effect that is consistent with an anomalous Hall contribution and is difficult to reconcile with a classical two-carrier description.}  
\end{abstract} 

\maketitle
Topological Kondo insulators (TKIs) such as SmB$_6$ have become model systems for exploring the interplay between strong electronic correlations and nontrivial band topology \cite{Dzero2016, Menth1969,Martin1979,Rosler2014}. In these materials, hybridization between localized $4f$ and itinerant conduction electrons opens a narrow bulk gap at the Fermi level, while symmetry-protected metallic surface states give rise to the characteristic low-temperature resistivity plateau \cite{Eo2019}. Following this description, a number of correlated narrow-gap semiconductors based on $3d$ electrons have drawn attention for exhibiting a similar dichotomy of bulk-insulating and surface-conducting behavior \cite{Petrovic2005, Xu2020, Changdar2020, Eo2023, Broyles2025}. Among these, FeSi and FeSb$_2$ stand out as nonmagnetic insulators with extraordinarily robust narrow bulk transport gaps and surface conduction channels at low temperature, prompting comparisons to SmB$_6$, despite their lack of heavy $f$-electrons \cite{Eo2023}.

FeSb$_2$, in particular, has attracted sustained interest as a narrow-gap correlated semiconductor with remarkable thermoelectric properties \cite{Sun2009, Zhao2011, Saleemi2016}. Its colossal thermopower at cryogenic temperatures has led to careful studies of its electronic structure which gave rise to its description as a correlated insulator \cite{Chikina2020, Li2024}. Spectroscopic measurements have identified fully gapped surface states and a thermally activated spin gap, suggesting a non-topological origin distinct from the Kondo-driven hybridization scenario of SmB$_6$ \cite{Li2024}. Despite recent theoretical models and experimental evidence, the microscopic origin of these surface states, and their relationship to the bulk electronic and magnetic structure, remains unresolved. 

The recent observations of magnetic correlations in the surface electrical transport of FeSb$_2$ \cite{Horn2026} has motivated the comparison to the magnetic surface states of FeSi \cite{Ohtsuka2021,Deng2023,Avers2024}, another $d$-electron system that has been similarly compared to TKIs \cite{Fang2018, Fu1994,Schlesinger1997,Breindel2023}. 
The evidence of surface magnetism in these materials motivates an interesting fundamental question of whether an intrinsic two-dimensional magnetically ordered surface state could be stable on the surface of a non-magnetic bulk material such as FeSi or FeSb$_2$. In two-dimensions, magnetic order is not expected for finite temperatures according to the Mermin-Wagner theorem \cite{Mermin1966} and has only recently been found in Van der Waals magnets with field induced magnetic anisotropy \cite{Gong2017}. To this end, the surface states of FeSi are a curious example of 2D magnetic order, which exists at zero field and finite temperature with minimal crystalline anisotropy \cite{Avers2024,Ohtsuka2021}. A theoretical description of the surface of FeSi is that of a spin-orbit coupled state formed from dangling bonds, which create a potential gradient normal to the surface associated with the Zak phase \cite{Ohtsuka2021,Zak1989}. As for FeSb$_2$, no such theoretical description has been claimed, however, magnetic correlations in the surface magnetotransport properties have been shown to exhibit large anisotropy \cite{Horn2026}, perhaps large enough to stabilize 2D magnetic order as has been argued for some Van der Waals magnets \cite{Gong2017,Xiao2022,AguileradelToro2023}. Notably, it has been predicted that in RuO$_2$, the bulk of which is non-magnetic \cite{Kiefer2025}, a 2D magnetic surface state may be present by a symmetry breaking [110]-surface reconstruction \cite{Ho2025}, similar to the mechanism used to describe the metallic surface states in FeSb$_2$ ARPES results \cite{Chikina2020}. Furthermore, similar to RuO$_2$, FeSb$_2$ is predicted to be an incipient altermagnet with magnetic order expected to be stable with chemical substitution \cite{Mazin2021} which has motivated recent interests in the electronic structure \cite{Phillips2025,Gauswami2025} and inducing magnetic order \cite{Bhandari2025}. {Reports of surface magnetism due to surface impurity concentrations induced by surface treatment in some systems exhibit an isotropic or amorphous magnetic surface state \cite{Coey2016}.}\color{black}

\begin{figure*}
    \centering
    \includegraphics[width=\textwidth]{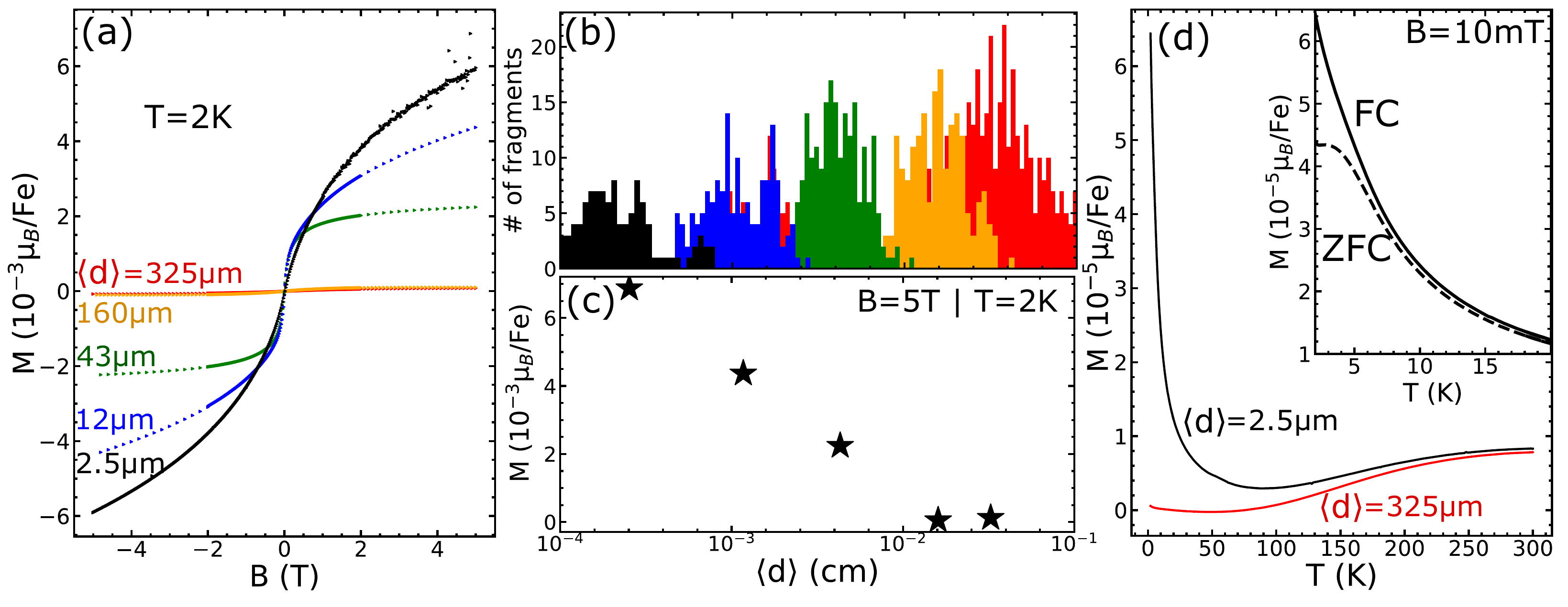}
    \caption{Magnetization scaling of the powder magnetic susceptibility with surface area (i.e. powder grain size). (a) Magnetization field dependence scaling with each powder step labeled by average diameter. (b) histograms of powder grain size for each powder grind step as calculated from powder images. (c) $B$ = 5 T, $T$ = 2 K magnetization for each average grain size. (d) $M$ vs $T$ at $B$ = 10 mT for the most coarse ($\langle d\rangle=325\mu m$) and most fine powders with inset for the finest ($\langle d\rangle=2.5\mu m$) showing a bifurcation between zero field cooled (ZFC) and field cooled (FC) data.}
    \label{fig:one}
\end{figure*}

In this work, we investigate the magnetic properties of the surface conducting states of FeSb$_2$, extracting the surface magnetic anisotropy and comparing to magnetotransport response. We characterize the surface contribution to the total magnetic susceptibility by measuring the surface area dependence of the magnetization of FeSb$_2$ powders obtained from systematically repeated grindings, finding that the surface magnetism retains the anisotropy of the bulk crystal structure and is correlated with magnetotransport.

Single crystals of FeSb$_2$ were synthesized using molten flux method as described in Ref.~\cite{Petrovic2005}. Powder fragments for magnetic susceptibility measurements were prepared with a clean ceramic crucible and grain size were measured using Zeiss smart-zoom optical microscope camera and calculated using python scipy toolkit to measure convex hull of individual grains (see Appendix~\ref{app:area}). Measurements of magnetization ($M$) and magnetic susceptibility ($\chi=\frac{M}{B}$, for small magnetic field, $B$) as a function of temperature ($T$) and magnetic field were collected using a Quantum Design MPMS-3 with a gel capsule and straw for powder and quartz rod for single crystal measurements. The gel capsule and straw used for the systematic surface area-dependent measurements were reused to preserve a nearly identical diamagnetic background. We started with a coarsely ground by mortal and pestle sample of FeSb$_2$, measured $M$ vs $T$, $M$ vs $B$ and captured the grain size distribution with microscope images, then further ground the sample and repeated measurements. 

The evolution of $M$ vs $B$, scaled by number of iron atoms determined by mass, increases with subsequent steps as demonstrated in Fig.~\ref{fig:one}a. Since each step creates smaller and smaller pieces of sample, the surface area per mass increases with each step and therefore Fig.~\ref{fig:one}a represents $M$ scaling with surface area. From analyzing the powder grain sizes from optical microscope images, we obtain the histogram representation of the average diameter of grains in each measurement step, shown in Fig.~\ref{fig:one}b, calculated from twice length of average distance from edge to center for each grain. {Since we did not use a sieve to separate grain sizes to achieve a more uniform size distribution, some non-uniformity is clear in Fig.~1b for example in the largest grain size sample ($\langle d\rangle=325\mu m$) which contains some portion of smaller grain sizes. Results from tracking $M$ at $T$ = 2 K and $B$ = 5 T in Fig.~\ref{fig:one}c shows a clear trend from $\langle d\rangle= 12\mu m$ to $\langle d\rangle=325\mu m$ in which the magnetization grows systematically with decreasing particle size.}\color{black} Fig.~\ref{fig:one}d compares the fine ($\langle d\rangle=2.5\mu m$) and coarse ($\langle d\rangle=325\mu m$) magnetization as a function of temperature, showing a bulk $M_{bulk}$ that is proportional to $T$ (consistent with a thermally activated moment \cite{Petrovic2005,Li2012}) and a surface moment ($M_{surface}$) that is inversely proportional to $T$ (consistent with free moments \cite{Mugiraneza2022}). The inset plot of Fig.~\ref{fig:one}d shows a bifurcation between zero-field cooling (ZFC) and field cooling (FC) data for the fine powder $M$, which {\color{black}indicates freezing or ordering of correlated moments associated with the surface of FeSb$_2$.}

\begin{figure}
    \centering
    \includegraphics[width=1\linewidth]{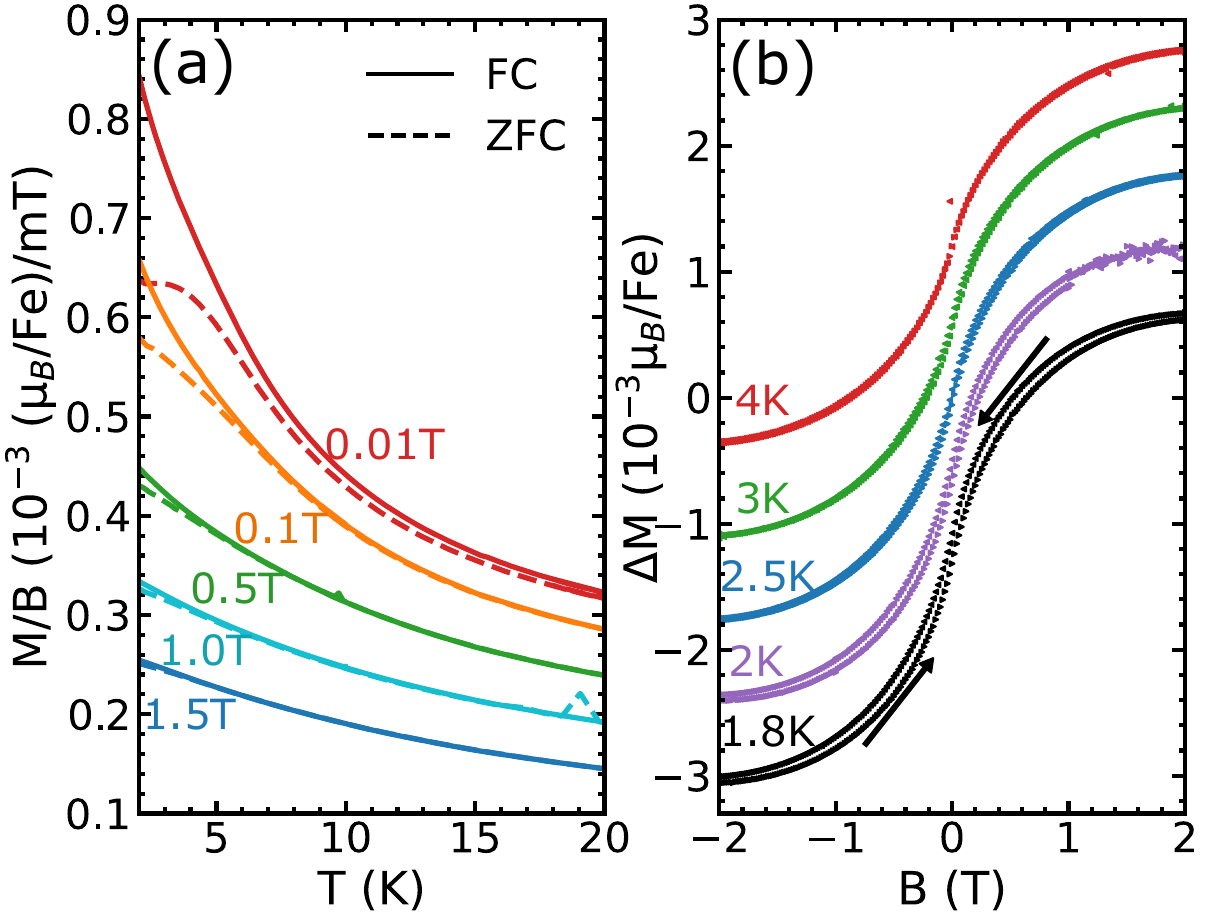}
    \caption{$M/B$ vs $T$ and magnetization with linear field dependence subtracted for clarity of hysteresis, $\Delta M$ vs $B$, is shown for fine FeSb$_2$ powder. These results show (a) field dependence of ZFC vs FC bifurcation and (b) temperature dependence of magnetic hysteresis.}
    \label{fig:two}
\end{figure}

Using the results from the fine powder, we are able to assume that the bulk magnetic susceptibility at low temperatures is sufficiently small that we can extract an effective surface magnetic moment ($\mu_{eff}$) using Curie-Weiss fit ($\chi = C/(T-\theta)$ for $\mu_{eff}$ $=\sqrt{8C}$). To estimate the total surface area, we use the fragments measured by Zeiss Smartzoom microscope as a representative sample and approximate the fragments by prolate spheroids with dimensions given by the minimum and maximum radius of individual fragments. {\color{black}Converting the Curie constant into a Curie constant per unit surface area (Appendix~\ref{app:area}) gives 97 $\mu_B$/nm$^2$ of spheroid surface area. Independently, the linearly additive magnetization at 2 K and 5 T gives 60 $\mu_B$/nm$^2$ (Table~\ref{table:magnetization}). These correspond to 11 and 7 $\mu_B$ per surface Fe, respectively, for a single unit-cell-thick surface layer. Both values are upper bounds, because the spheroid approximation underestimates the true area of irregular, rough fragments: for ground crystalline powders, surface areas measured by Brunauer--Emmett--Teller (BET) gas adsorption exceed the geometric value by a factor $r\approx7$ (range 4--11) \cite{Brantley2000}. For $r\approx7$, both estimates fall to 1--4 $\mu_B$ per surface Fe, which is physically reasonable (Fig.~\ref{fig:S5}; details in Appendix~\ref{app:roughness}).} The comparison between FeSb$_2$ powder, FeSi powder and FeSi thin films are given by table \ref{table:magnetization}. {\color{black}The remaining difference between the powder-derived and thin-film moments is within the surface-area uncertainty of the powder measurements.}

\begin{table}
\caption{Comparison of surface magnetic moment between FeSb$_2$ and FeSi.}
\label{table:magnetization}
{
\centering
\begin{tabular}{p{0.1\textwidth} p{0.12\textwidth} p{0.08\textwidth} p{0.15\textwidth} }
\hline\hline
material  &  M ($\mu_{B}/$nm$^2$) & geometry & method \\
\hline
 FeSb$_2$ & 60 & powder & magnetization \\
 FeSb$_2$ & 97 & powder & Curie constant \\
 FeSi \cite{Avers2024} & 500  & powder & magnetization \\
 FeSi \cite{Ohtsuka2021} & 22.5  & thin film & magnetization \\

\hline\hline
\end{tabular}\par
}
\end{table}

Using the $\langle d\rangle=2.5\mu m$ powder step, we examined the bifurcation from Fig.~\ref{fig:one}d as a function of magnetic field as shown in Fig.~\ref{fig:two}a. Traces of a bifurcation between ZFC and FC data disappear between 1 T and 1.5 T field.  This is consistent with the results from low temperature $M$ vs $B$ sweeps shown in Fig.~\ref{fig:two}b, which shows $\Delta M$ vs $B$, where $\Delta M$ is the result from subtracting a linear slope from high field data in order to better show the small hysteresis, which closes below 2 T at 1.8 K. The shape of the magnetic hysteresis is unusual for a single phase magnetically ordered system and appears to be ``wasp-waisted" (i.e. the hysteresis is narrow at low fields and wider at higher fields), which may be attributed to large magnetocrystalline anisotropy, multiple magnetic phases, or coexisting superparamagnetism \cite{Roberts1995,Magno_de_Lima_Alves2017}.  While the existence of a bifurcation in our data alone does not indicate ferromagnetism, as some spin-glass systems show similar behavior \cite{Binder1986}, {\color{black}we take the field scaling of the bifurcation shown in Fig.~\ref{fig:two}a along with the hysteresis shown in Fig.~\ref{fig:two}b as evidence of collective freezing or ordering of the surface moments, although superparamagnetic blocking of clusters or cluster-glass freezing could produce similar features.} We also present in Appendix~\ref{app:ac} some AC susceptibility up to 400 Hz which, while we cannot rule out a spin-glass completely, puts a limit on the time scale of spin dynamics in a spin glass interpretation of the surface magnetism. \color{black}

\begin{figure}
    \centering
    \includegraphics[width=0.7\linewidth]{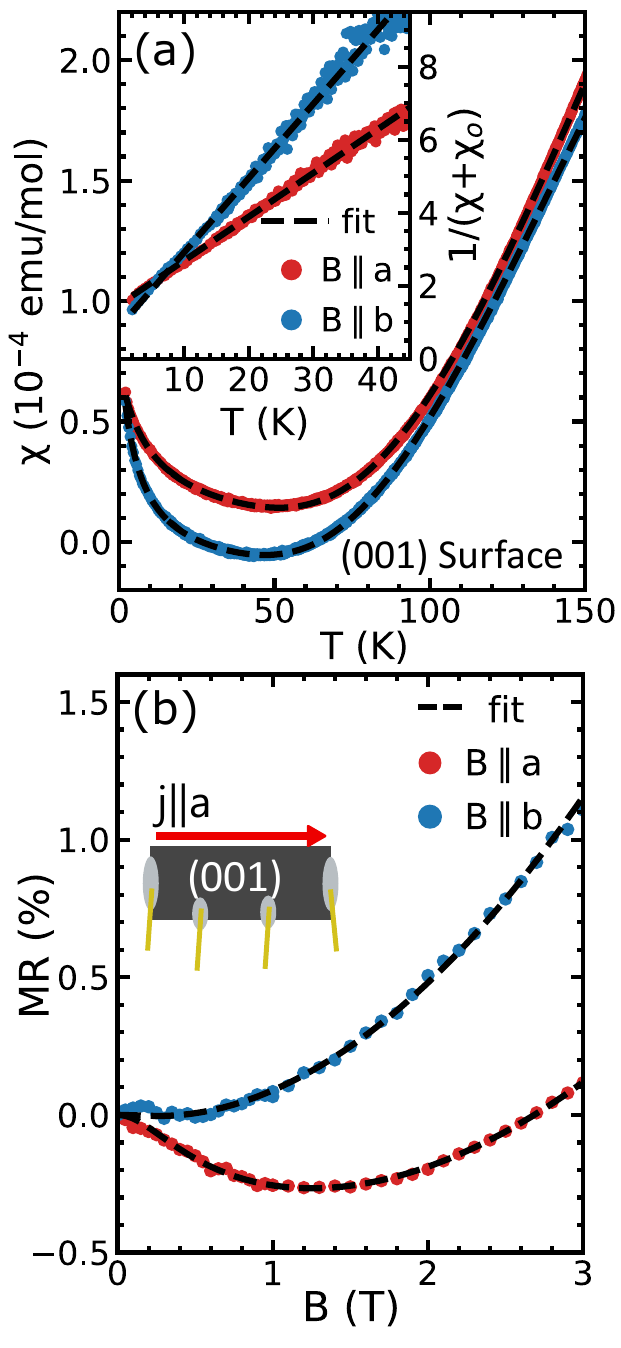}
    \caption{Anisotropy of surface magnetization compared with surface magnetotransport. (a) magnetic susceptibility of a thin, primarily (001) surface, rough polished single crystal for field along $a$ and $b$-axes. Dashed line represents a fit, which is a linear combination of bulk spin-gap susceptibility and surface Curie-Weiss susceptibility. (b) Magnetoresistance ($MR=100\%*(R(B)-R(0))/R(0)$) of a thin, primarily (001) surface sample with current along $c$-axis and magnetic field along $a$ (red) and $b$ (blue) axes. Dashed line represents fit to linear combination of hopping conduction (positive MR) and magnetic scattering (negative MR).}
    \label{fig:three}
\end{figure}

From comparing the coarse (bulk-like) and fine (surface-like) powder magnetization in Fig.~\ref{fig:one}, it is clear that the Curie-Weiss $M$ vs $T$ at low temperature may be attributed to the surface magnetism ($\chi_{surface} = \chi_0 + C/(T-\theta)$) and the thermally activated temperature dependent magnetization (increasing with temperature) may be attributed to the bulk magnetism. The bulk magnetization has been described as a spin-gap or Kondo-like behavior \cite{Petrovic2005}, with the following expression for the temperature dependence:

\begin{equation}
    \chi_{bulk} = \frac{-2N\mu_B^2(1-exp(\beta W))}{W(1+exp(\beta W))(1+exp(\beta (\Delta + W)))}    \label{eqn:spin-gap},
\end{equation}

where $N$ is the number of states, $W$ is the width of density of states bands, which are separated by energy $2\Delta$ and $\beta=1/kT$. At low temperatures, this gives a vanishingly small moment that is minimally temperature dependent, which is associated with the singlet ground state of the bulk. In this limit, the surface Curie-Weiss behavior is the leading behavior, which is consistent with Fig.~\ref{fig:one}d and Fig.~\ref{fig:three}a. 

{\color{black}The surface-area-dependent moment could instead arise from grinding-induced defects, a different Fe valence state, or an oxide layer \cite{Sundaresan2006,Coey2016}, and its magnitude alone does not exclude this.  However, such oxides order far above room temperature, whereas the fine and coarse powders have the same magnetization at 300 K (Fig.~\ref{fig:one}d). Without surface-sensitive chemical probes (e.g., XPS), we cannot fully exclude a chemically modified near-surface layer as the host of the moments. The magnetic anisotropy offers a further test, as dilute Fe$^{3+}$ moments or an amorphous defect layer are expected to be nearly isotropic, whereas moments tied to the crystalline surface should reflect the anisotropy of the lattice.}

In order to compare the surface magnetization with the observed anisotropic magnetic correlations in the surface magnetotransport \cite{Horn2026}, we prepared a thin rough (001)-oriented surface (such that field is measured in $a$-$b$ plane) single-crystal of FeSb$_2$, less than 0.1 mm thick with silhouette area of 6 mm$^2$, which allows us to maximize surface area while maintaining crystal orientation. Since the ratio of surface area to bulk volume is not as large as for a fine powder, the low temperature surface moment is small relative to the high temperature bulk magnetic moment. However, at temperatures below 50 K, the bulk moment is primarily singlet-like and is small relative to the surface magnetic moment. This is clearly shown in the inset of Fig.~\ref{fig:three}a, which shows a clear Curie-Weiss behavior for which $1/(\chi+\chi_0)$ is linear in $T$ for some $\chi_0$ as shown by good agreement in black dash lined linear fit. Furthermore, we were able to fit the full range of susceptibility to a sum of a Curie-Weiss surface magnetization and a thermally activated bulk magnetization (equation \ref{eqn:spin-gap}) as shown by the dashed lines in the main plot of Fig.\ref{fig:three}a. In comparing the surface Curie-Weiss behavior for the same (001) surface sample for magnetic field along $a$ and $b$ crystallographic axes, there is a clear magnetic anisotropy present in the low temperature magnetization. Results from Curie-Weiss fits at low temperature (first two results columns of table \ref{table:Anisotropy}) show that the Curie-Weiss effective moment for field along a is approx 30\% larger than for field along $b$ along with a Weiss temperature of 12 K for field along $a$-axis and 4 K for field along $b$-axis. This comes together to imply a magnetocrystalline anisotropy for which the  $a$-axis is the magnetic easy axis with ferromagnetic correlations. 

To compare the surface magnetic anisotropy to the magnetic correlations in the surface magnetotransport properties of the FeSb$_2$ (001) surface, we prepared a transport sample with current along $a$-axis and the magnetic field rotating in the $a-b$ plane. In a recent study, we demonstrated that the surface magnetotransport properties of FeSb$_2$ are independent of the relative alignment of the current and magnetic field \cite{Horn2026}. Therefore, in this experiment we should be sensitive only to the magnetocrystalline anisotropy. Instead, the magnetoresistance (MR$(B)=100\%*\frac{R(B)-R(0)}{R(0)}$) features a positive MR component due to the effect of the magnetic field on the hopping conduction ($\propto \exp(B)$) and a negative MR component due to the effect of the magnetic field on suppressing magnetic scattering ($\propto -\log(1+(B/B_0)^2)$). This comes together to give a total magnetoresistance:

\begin{equation}
    \label{eqn:Magnetoresistance}
    \frac{R(B)-R(0)}{R(0)} = -a\log(1+b^2B^2) + \exp(c^2B^2)-1
\end{equation}

The first fitting parameter which relates to the amplitude of the negative MR term, $a = A_1 J\rho_F$, describes the Kondo coupling between local moments and charge carriers in which $A_1$ is a numerical prefactor, $J$ is the strength of the exchange interaction and $\rho_F$ is the density of states at the Fermi energy. The second fitting parameter which relates to the field scale of the negative MR term, $b=\frac{g_0\mu}{\alpha k_BT}$, describes the local moments involved in the scattering process in which $g_0$ is the Lande g-factor, $\mu_B$ is the Bohr magneton, $k_B$ is the Boltzmann constant and $\alpha$ is an empirical broadening parameter. The last fitting parameter which relates to the size and field scale of the positive MR term, $c^2=\frac{te^2a_{hop}}{\hbar^2N}\left(\frac{T_{VRH}}{T}\right)^{3/4}$, describes the influence of the magnetic field on the localization of hopping sites in which $a_{hop}$ is the hopping conduction length-scale, $N$ is the hopping site density, t is a numerical constant that is derived from percolation theory, and $T_{VRH}$ is the hopping conduction temperature scale which is set by the effective size and density of states of hopping sites. \cite{Khosla1970,Shklovskii1984} \color{black}

\begin{table}
\caption{Summary of the magnetic anisotropy from fits to magnetic susceptibility ($\mu_{eff}$ and $\theta$) and magnetoresistance ($\mu g_0/\alpha$ and $A_1J\rho_F$) .}
{
\centering
\begin{tabular}{p{0.05\textwidth} p{0.12\textwidth} p{0.07\textwidth} p{0.12\textwidth} p{0.08\textwidth}}
\hline\hline
axes  &  $\mu_{eff}$ ($\mu_{B}$/Fe) & $\theta$ (K) & $\mu g_0/\alpha$ (eV/T) & $A_1J\rho_F$ \\
\hline
 $a$ & 8.16$\times10^{-2}$ & 12.4 & 5.78$\times10^{-4}$ & 0.134\\
 $b$ & 6.27$\times10^{-2}$ & 4.30 & 4.84$\times10^{-4}$ & 0.028\\

\hline\hline
\end{tabular}\par
}
\label{table:Anisotropy}
\end{table}

The (001) surface MR at 2 K fits well to the combined hopping conduction and magnetic scattering model, as evidence from the black dashed fit line shown in Fig.~\ref{fig:three}b. For field along the b-axis, there is no evidence of magnetic scattering suppression (i.e. MR vs $B$ is monotonically increasing with field), whereas for field along the $a$-axis, the low field ($B<$2 T) is dominated by the negative magnetoresistance contribution from suppressing magnetic scattering. This surface magnetotransport anisotropy is qualitatively consistent with the surface magnetization reported in Fig.~\ref{fig:three}a. Table \ref{table:Anisotropy} compares the results from fitting the magnetoresistance and magnetization data for fields along $a$ and $b$ axes. The relative anisotropy obtained for $\mu$ (assuming isotropic $g_0/\alpha$) is consistent across these fits. Similarly, the anisotropy in the magnetotransport parameter that describes the coupling between charge carriers and local moments (i.e. $A_1J\rho_F$) is similar to the anisotropy in the Weiss temperature, which describes the coupling strength between moments. {\color{black}Quantitatively, the $a$/$b$ ratio of the Curie-Weiss moment (1.30) and that of the magnetic scattering field scale in the magnetoresistance (1.19) agree to within 10\%, although they come from a thermodynamic and a transport measurement with different sensitivities to bulk and surface. This is consistent with the moment for the Curie-Weiss susceptibility being the same moments responsible for scattering in the surface conduction channel. While this does not by itself prove that the moments reside within the conducting surface state rather than in a structurally modified near-surface layer, it constrains any such layer to be crystallographically coherent and electronically coupled to the surface conduction channel.}

\begin{figure}
    \centering
    \includegraphics[width=0.8\linewidth]{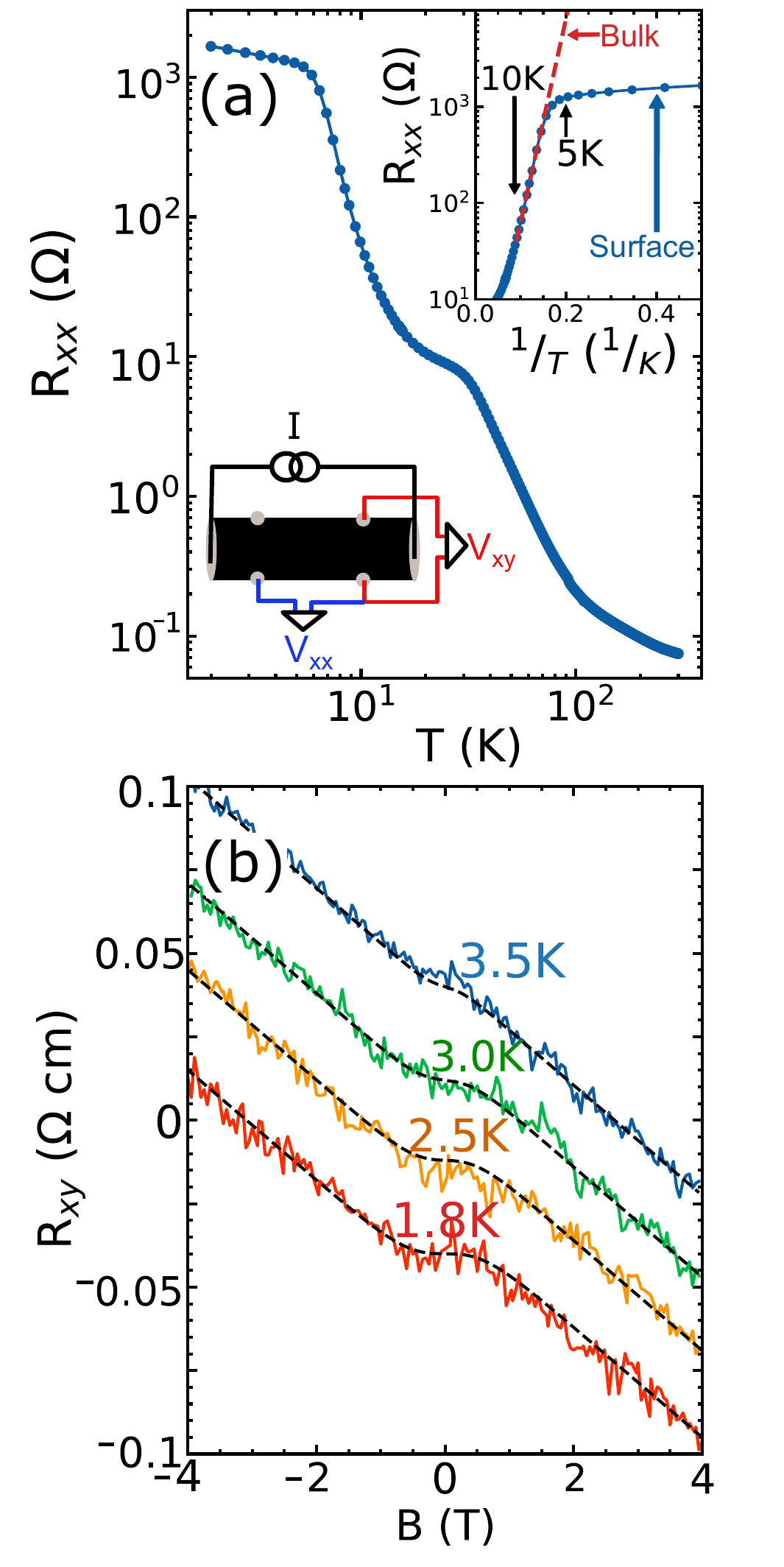}
    \caption{Hall effect {\color{black}showing a small nonlinear component} in the surface dominated transport regime. (a) resistance vs temperature of primarily (101) surface with current along $b$-axis. (a, inset) inverse temperature dependence of resistance showing the crossover from primarily bulk electrical conduction around 10 K to primarily surface conduction below 5 K. (b) Hall effect in the primarily surface conduction regime. Dashed line represents linear combination of single channel Hall and {\color{black}a saturating ($\tanh$) component, consistent with an anomalous Hall contribution (see text and Fig.~\ref{fig:S6} for a constrained two-carrier analysis).}
    }
    \label{fig:four}
\end{figure}

Finally, to complete the study, we examined the low field Hall effect behavior to look for evidence of an anomalous contribution, as reported in FeSi \cite{Avers2024,Ohtsuka2021}. Understanding the Hall effect in semiconductors with surface states is notoriously challenging due to the strong dependence on surface quality with subsurface cracks from sample preparation playing a large role in the surface transport \cite{Eo2020}. To achieve the best surface quality, we used an as-grown facet of FeSb$_2$, for which the (101) surface is the best candidate as this direction forms large clean surfaces from synthesis. Fig.~\ref{fig:four} shows the resistance (Fig.~\ref{fig:four}a) and Hall effect (Fig.~\ref{fig:four}b) results for electrical transport on the (101) surface (field along [101] direction) with current along $b$-axis. The resistance as a function of temperature from Fig.~\ref{fig:four}a is used to determine the crossover from primarily-bulk to primarily-surface transport regime between 5 K and 10 K. It was previously shown that for this material, the bulk transport properties is characterized by an extraordinarily robust activated behavior, showing no signs of impurity conduction, therefore we assume that the bulk resistance at low temperatures can be predicted by a linear fit to inverse temperature (i.e. $R(T)\propto \exp(T_o/T)$) \cite{Eo2023}. The inset of Fig.~\ref{fig:four}a shows that, using this assumption, at 5 K the bulk of FeSb$_2$ is several times more resistive that the surface and by 3.5 K, conduction is near entirely through the surface transport channel. Fig.~\ref{fig:four}b shows that, in the surface transport regime, the Hall effect shows a low field curvature, which increases with decreasing temperature below 3.5 K. A dashed line fit is used to compare the Hall effect data with a linear and Langevin function combination as to compare with the surface magnetization field dependence for a ferromagnetic contribution. Unfortunately, the Hall mobility is low for the surfaces we've been able to study and therefore this anomalous response is too faint to be able to discern any hysteresis that would be expected for an anomalous Hall effect.

{\color{black}A multi-carrier scenario can also give a nonlinear Hall effect, so we performed a constrained two-carrier analysis using the measured sheet conductance. Reproducing the curvature requires a low-density, high-mobility ($\sim$1 m$^2$/Vs) minority channel. Such a channel could be consistent with bulk FeSb$_2$ carriers \cite{Takahashi2011}, but its conductance is nearly temperature independent between 1.8 and 3.5 K. Bulk conduction in FeSb$_2$, by contrast, is robustly activated with no sign of impurity conduction \cite{Eo2023}, and must fall by more than eight orders of magnitude over this range, which rules out a bulk origin. The fit model and results are given in Appendix~\ref{app:hall} (Fig.~\ref{fig:S6}). Including an anomalous component to the ordinary Hall effect, on the other hand, fits the data equally well. Its amplitude grows on cooling, and its saturation field (0.2--1 T) is comparable to the field scale over which the ZFC/FC bifurcation in the magnetic susceptibility is suppressed (Fig.~\ref{fig:two}a). However, hysteresis is not resolved, therefore we describe this nonlinear response as consistent with, but not conclusive evidence for, an anomalous Hall contribution in the surface transport of FeSb$_2$.}

In summary, we report on {\color{black}a study} of the magnetic properties of the low temperature surface states on FeSb$_2$. Through systematic study of the magnetization with increasing surface area, we were able to ascribe the observed low temperature Curie-Weiss moment in pure FeSb$_2$ to {\color{black}moments associated with the surface, with a magnitude of order 1--3 $\mu_B$ per surface Fe once realistic surface areas are taken into account. We also find evidence that these moments freeze or order collectively,} with a FC/ZFC bifurcation closing around 15 K at 10 mT and magnetic hysteresis. Furthermore, we measure the susceptibility of a flat, high surface-area-to-bulk single crystal for field along $a$ and $b$ crystal axes to find that this surface magnetism retains the anisotropy of the bulk crystal structure, which {\color{black}is difficult to reconcile with} an isotropic or polycrystalline surface state, which might be expected for a disordered magnetic oxide layer formed on the surface. Additionally, we compare the magnetotransport anisotropy on the same surface and field directions and find good agreement in the anisotropy of the transport and magnetic fit parameters between field along $a$- and $b$-axis. The emergence of an anisotropic magnetic surface state in an otherwise nonmagnetic insulator suggests that, although the bulk of FeSb$_2$ electronic structure is robust against impurities, it remains sensitive to the local symmetry breaking and perturbations at its surfaces. The resulting surface magnetism, which couples strongly to the crystal symmetry, indicates that the nonmagnetic ground state of FeSb$_2$ lies in close proximity to a magnetic instability.

While these results establish a surface area dependent and anisotropic magnetic response in FeSb$_2$, the microscopic origin of this state remains to be determined. A theoretical framework on how surface symmetry breaking, reconstruction, reduced coordination, or band bending modify the narrow-gap electronic structure and whether these perturbations can destabilize the nonmagnetic bulk state in favor of surface-localized moments still remain to be addressed. Additionally, direct surface sensitive probes such as x-ray magnetic circular dichroism (XMCD), magneto-optical Kerr effect (MOKE), or magnetic force microscopy (MFM) would be valuable for testing the spatial localization, symmetry, and possible domain structure of the magnetic response. Such studies will be important for distinguishing an intrinsic surface magnetic state from disorder-induced effects. {\color{black}In particular, surface-sensitive chemical characterization (e.g., XPS) of crystals and powders before and after grinding, a direct BET determination of the powder surface area, and measurements correlating the magnetic response with controlled surface preparation and independently measured surface conductance would directly test whether the moments reside in the conducting surface channel or in a chemically modified near-surface layer.}
\color{black}

\begin{acknowledgments}
Research at the University of Maryland was supported by
the Gordon and Betty Moore Foundation’s EPiQS Initiative Grant No. GBMF9071 (materials synthesis),
the U.S. National Science Foundation Grant No. DMR2303090 (sample preparation), 
the Air Force Office of Scientific Research Grant No. FA9950-22-1-0023 (experimental measurements), 
the NIST Center for Neutron Research, and the Maryland Quantum Materials Center. 
\end{acknowledgments} 

\appendix

\section{Powder surface area and Curie constant per unit area}
\label{app:area}

\begin{figure}
    \centering
    \includegraphics[width=1\linewidth]{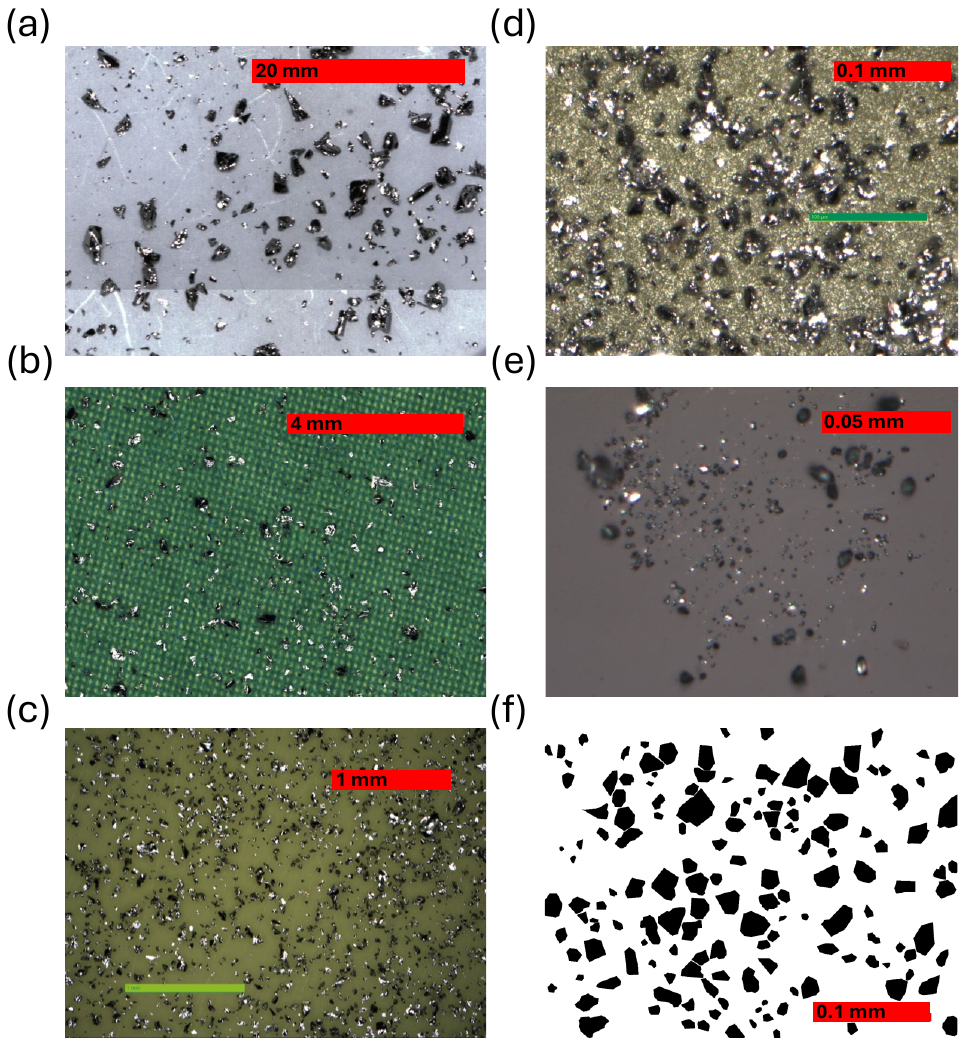}
    \caption{Microscope images of the powder grains using the built-in stitching tool from Zeiss smartzoom microscope (a-e) and an example silhouette used in convex hull analysis from $\langle d \rangle$=12$\mu$m sample.}
    \label{fig:S2}
\end{figure}

\begin{figure}
    \centering
    \includegraphics[width=0.75\linewidth]{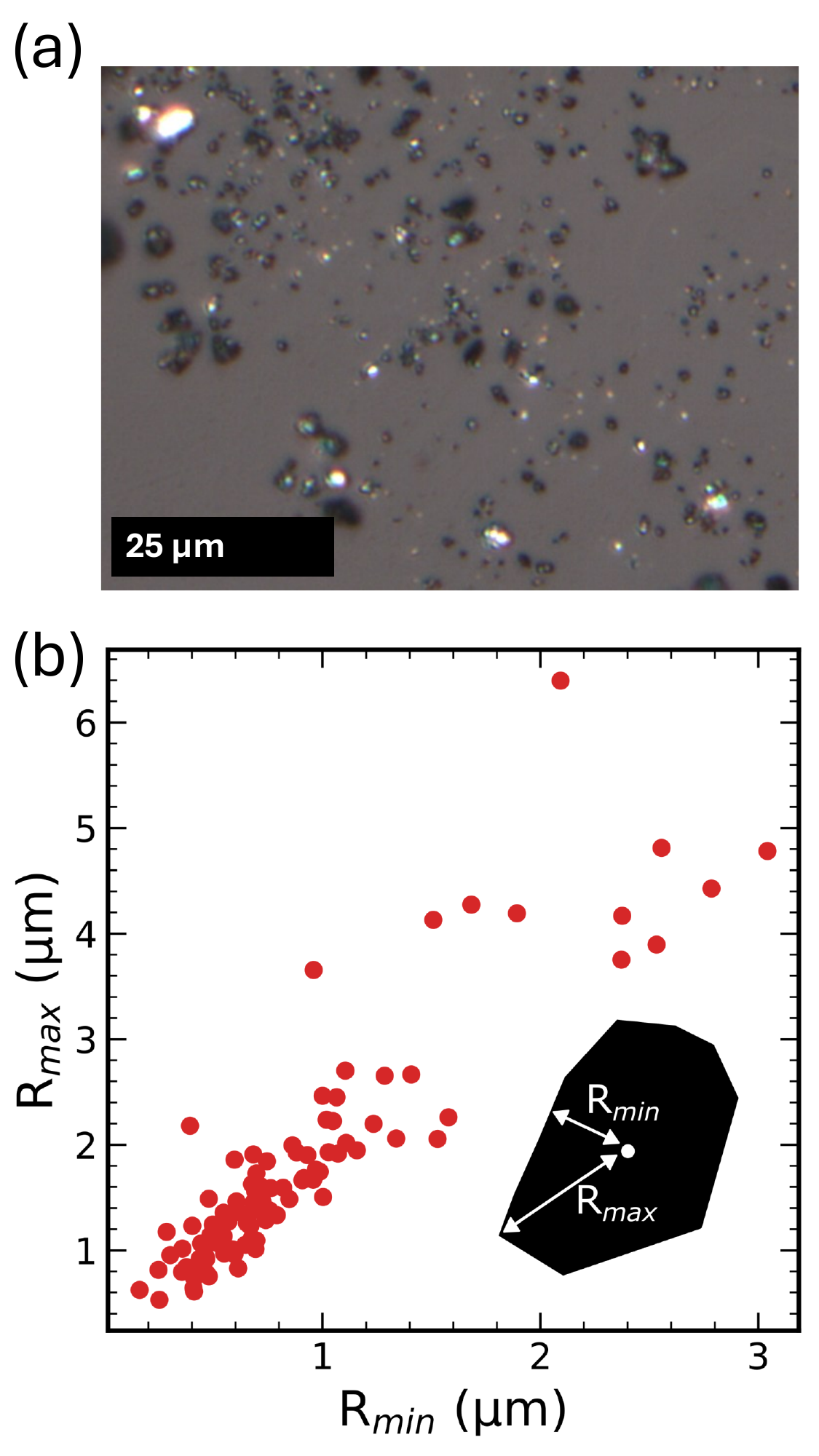}
    \caption{(a) Enlarged photo of $\langle d \rangle$=2.5$\mu$m fragments along with (b) the dimensions of grain size silhouettes calculated from convex hull analysis.}
    \label{fig:S1}
\end{figure}

To estimate the surface area of the powder fragments, we used a Zeiss smartzoom microscope to take scaled images of the poweder and analyze the images to extract out a histogram of fragment shapes. To convert the 2D shapes into 3D surface areas and volumes, we approximated the fragments as prolate spheroids characterized by a minimum and maximum radii of $R_{min}$ and $R_{max}$, respectively. These were obtained by fitting the silhouette images of individual fragment to an outline using convex hull analysis in python. Results from the $\langle d \rangle$=2.5$\mu$m sample are shown in Fig.~\ref{fig:S1}. Photos of each step of smashing the powder is shown in Fig.~\ref{fig:S2}a-e and an example silhouette image from which convex hull was calculated is shown in Fig.~\ref{fig:S2}f.

For each fragment, an approximate prolate spheroid is calculated using the silhouette data for $R_{min}$ and $R_{max}$. We estimate the volume as

\begin{equation}
    V = \frac{4}{3}\pi R_{min}^2 R_{max}
    \label{eqn:V}
\end{equation}

and the surface area as

\begin{equation}
    S = 2\pi R_{min}^2 + \frac{2\pi}{e} R_{min}R_{max}\sin^{-1}(e)
    \label{eqn:S}
\end{equation}

where $e=\sqrt{\frac{R_{max}^2-R_{min}^2}{R_{max}^2}}$ is the eccentricity of the approximate prolate spheroid. Assuming the collection of fragments samples in the set of silhouettes that were measured is representative of the entire sample of fragments, we use the net surface area and volume ratio and the volume obtained from the mass of the sample ($m$ = 0.0313 g for the $\langle d \rangle$=2.5$\mu$m sample) and density ($\rho$ = 8.17 g/cm$^{-3}$ from crystal data \cite{FeSb2_Springer}). The net estimated surface area is therefore given by

\begin{equation}
    S_{net} = \left(\frac{\sum_i S_i}{\sum_j V_j}\right) \frac{m}{\rho}
    \label{eqn:Snet}
\end{equation}

where $V_j$ and $S_i$ represent independent sums on volumes and surface areas using equations \ref{eqn:V} and \ref{eqn:S}, respectively.

{\color{black}The low-temperature susceptibility of the finest powder (Fig.~\ref{fig:one}d, normalized to the total number of Fe atoms in the sample, as in all panels of Fig.~\ref{fig:one}) follows a Curie law between 5 and 20 K, with a Curie constant $C\approx1.3\times10^{-2}$ emu\,K/mol-Fe, i.e.\ an effective moment $\mu_{eff}=\sqrt{8C}\approx0.33\,\mu_B$ per Fe atom in the sample \cite{Mugiraneza2022}. Assuming the bulk contribution to the low-temperature Curie constant is negligible, this can be converted to a Curie constant per unit surface area using the expression from equation \ref{eqn:Snet}. Because the Curie constant is proportional to the sum of the \emph{squared} moments, $C\propto\sum_i\mu_i^2$, it is $\mu_{eff}^2$, not $\mu_{eff}$, that scales with the number of contributing atoms:
\begin{equation}
    C_A \equiv \mu_{eff}^2 \left(\frac{\#Fe/u.c.}{V_{u.c.}}\right) \left(\frac{\sum_j V_j}{\sum_i S_i}\right),
    \label{eqn:CA}
\end{equation}
where $\# Fe/u.c.$=2 \cite{Du2021}, $V_{u.c.}$=0.125 nm$^3$ \cite{FeSb2_Springer}, and $\frac{\sum_j V_j}{\sum_i S_i}$ = 632.5 nm, giving $C_A\approx1.1\times10^{3}\,\mu_B^2$/nm$^2$. The Curie constant alone determines only the product of the number of moments and the square of each moment. If the moments reside on $N_A$ sites per nm$^2$, each carries
\begin{equation}
    \mu_{loc}=\sqrt{\frac{C_A}{N_A}},
    \label{eqn:muloc}
\end{equation}
and the 2D moment density is $N_A\,\mu_{loc}=\sqrt{N_A\,C_A}$. Taking every Fe atom in the outermost unit-cell layer as a moment-bearing site ($N_A=\sigma_{Fe}\approx8.6$ nm$^{-2}$, see Appendix~\ref{app:roughness}) gives $\mu_{loc}\approx11\,\mu_B$ per surface Fe and $\approx97\,\mu_B$/nm$^2$ of spheroid area. A larger true surface area, $S_{true}=r\,S_{spheroid}$, reduces both by a factor of $\sqrt{r}$: for $r\approx7$, $\mu_{loc}\approx4\,\mu_B$ per surface Fe. The magnetization, which is linearly additive and requires no Curie-Weiss model, is analyzed in Appendix~\ref{app:roughness} and gives a consistent result.}

\section{AC susceptibility}
\label{app:ac}

\begin{figure}
    \centering
    \includegraphics[width=\linewidth]{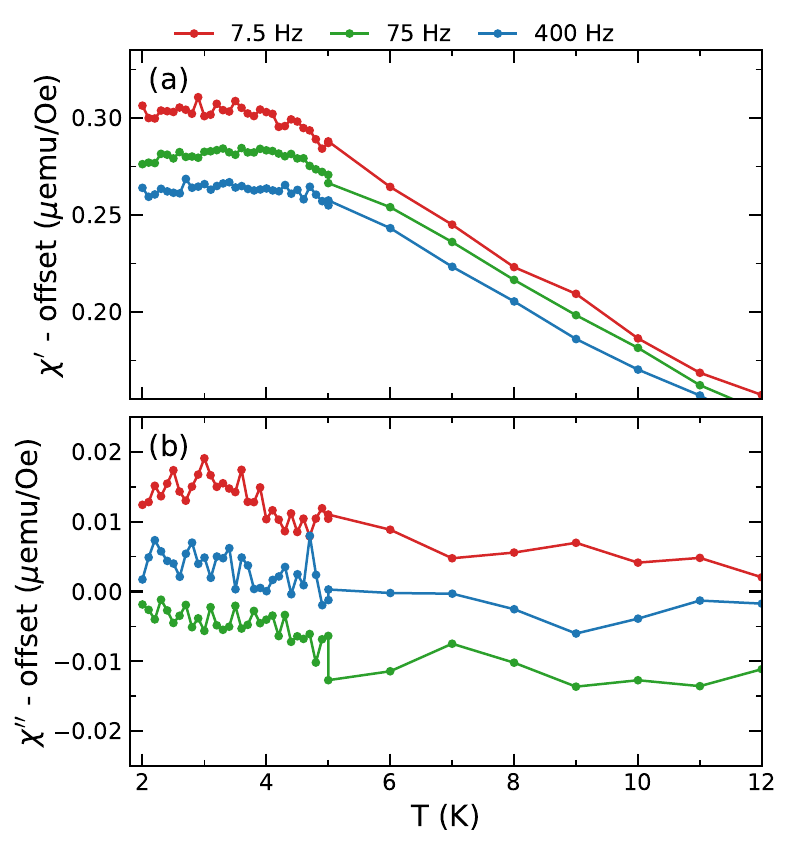}
    \caption{AC susceptibility data for fine powder FeSb$_2$ at B = 0.01 T applied field.}
    \label{fig:S4}
\end{figure}

To further evaluate the nature of the surface magnetism, i.e. whether the bifurcation in the susceptibility can be attributed to long-range order or spin-glass physics, we've measured the AC susceptibility of FeSb$_2$ powder as shown in Fig.~\ref{fig:S4}. The absence of a clear frequency dependence up to 400 Hz is evidence against a spin-glass surface state with slow time dynamics (see e.g. Binder and Young, Rev. Mod. Phys. 1986 \cite{Binder1986}). While we concede that these results along do not rule out completely the absense of spin-glass transition at the surface of FeSb$_2$, we argue that this, along with the arguments put forth above, point to a long range ordered surface state.

{\color{black}
\section{Surface-area uncertainty and moment per surface Fe}
\label{app:roughness}

\begin{figure}
    \centering
    \includegraphics[width=0.85\linewidth]{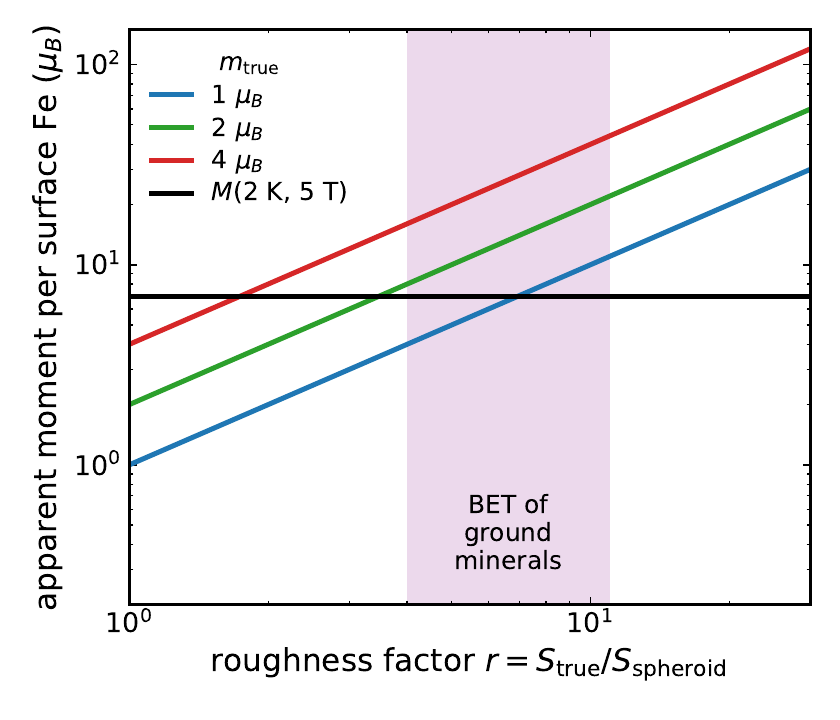}
    \caption{\color{black}Apparent moment per surface Fe inferred from the magnetization using the prolate-spheroid surface area, as a function of the roughness factor $r=S_{true}/S_{spheroid}$, for fixed true moments of 1, 2 and 4 $\mu_B$ per Fe in a single unit-cell-thick surface layer (colored lines). The horizontal line is the measured fine-powder surface magnetization at 2 K and 5 T. Shaded: ratio of Brunauer--Emmett--Teller (BET) gas-adsorption to geometric surface area for laboratory-ground crystalline powders \cite{Brantley2000}.}
    \label{fig:S5}
\end{figure}

The magnetization is linearly additive, so the surface magnetization per unit area is
\begin{equation}
    M_A = M_{surf}\, n_{Fe} \left(\frac{V}{S}\right),
    \label{eqn:MA}
\end{equation}
where $M_{surf}\approx5.9\times10^{-3}\,\mu_B$/Fe is the 2 K, 5 T magnetization of the finest powder minus that of the coarsest powder, $n_{Fe}=16$ nm$^{-3}$, and $V/S=632.5$ nm. This gives $M_A\approx60\,\mu_B$/nm$^2$ of spheroid area. The Fe areal densities of a unit-cell-thick layer are 5.3, 9.6 and 10.7 nm$^{-2}$ for the (001), (100) and (010) faces (average $\sigma_{Fe}\approx8.6$ nm$^{-2}$). Only a fraction
\begin{equation}
    f_0 = \frac{\sigma_{Fe}}{n_{Fe}\,(V/S)} \approx 8.5\times10^{-4}
    \label{eqn:f0}
\end{equation}
of all Fe atoms therefore lie in the outermost unit-cell layer, and the apparent moment is $M_{surf}/f_0\approx7\,\mu_B$ per surface Fe.

This number is an upper bound with respect to the surface area, because the true area $S_{true}=r\,S_{spheroid}$ exceeds the spheroid estimate for three reasons.

(i) \emph{Fragment shape.} We generated angular fragments as convex hulls of random points within triaxial ellipsoids, placed each on its flattest face, and analyzed its silhouette exactly as for the measured grains ($R_{min}$, $R_{max}$ from the centroid-to-edge distances). For typical fragment shapes, the true specific surface exceeds the spheroid estimate by a factor of 1.2--1.7. This is a lower bound, since the convex hull ignores steps and cracks.

(ii) \emph{Sub-resolution fine particles} Only fragments large enough to be identified in the microscope could be accounted for in our surface area measurements. For example, if 2\% of the total powder mass is comprised of 200 nm diameter particles, the surface area will be larger by a factor of 1.4. 

(iii) \emph{Microroughness.} For laboratory-ground crystalline powders, surface areas measured by Brunauer--Emmett--Teller (BET) gas adsorption exceed the sphere-equivalent area by a factor of 4--11 \cite{Brantley2000}.

As shown in Fig.~\ref{fig:S5}, a microroughness factor of $r\approx4$--11 brings the moment to 1--3.5 $\mu_B$ per surface Fe. 

\section{Constrained two-carrier analysis of the Hall effect}
\label{app:hall}

\begin{figure*}
    \centering
    \includegraphics[width=\linewidth]{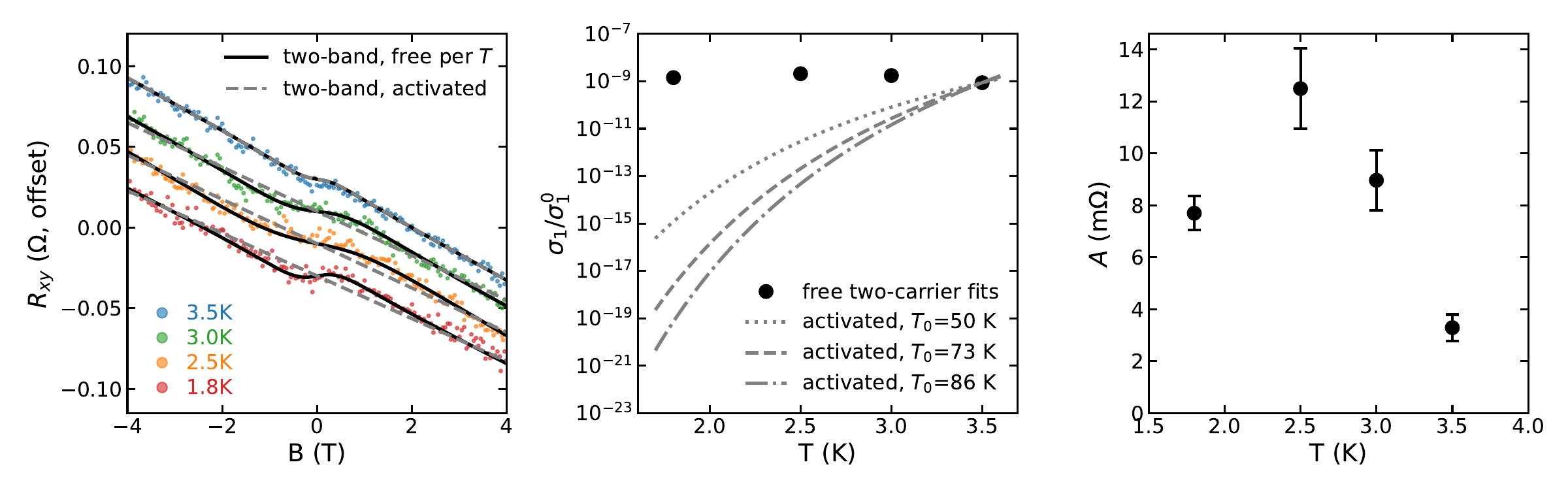}
    \caption{\color{black}(a) Antisymmetrized Hall resistance (offset for clarity) with the best two-carrier fits at fixed sheet conductance, with free parameters at each temperature (solid) and with the second channel constrained to follow thermally activated behavior ($T_0$ = 73 K, the bulk Hall activation temperature of Ref.~\cite{Eo2023}; dashed). (b) Sheet conductance of the minority channel required by the free two-carrier fits (symbols), compared with the activated temperature dependence of Eq.~(\ref{eqn:activated}) expected for bulk carriers (lines), for $T_0$ = 50 K (slope of the bulk line in Fig.~\ref{fig:four}a), 73 K (bulk Hall coefficient) and 86 K (bulk Corbino resistivity) \cite{Eo2023}. All curves are matched to the fitted value at 3.5 K, and $\sigma_1^0$ is the prefactor of the $T_0$ = 73 K curve. (c) Amplitude $A$ of the saturating component in $R_{xy}=R_0B+A\tanh(B/B_s)$.}
    \label{fig:S6}
\end{figure*}

For two carrier species with sheet conductances $\sigma_i=e n_i|\mu_i|$ and signed mobilities $\mu_i$ ($i=1,2$), the Drude conductivity tensor is
\begin{equation}
    \sigma_{xx}=\sum_{i=1,2}\frac{\sigma_i}{1+\mu_i^2B^2},\qquad
    \sigma_{xy}=\sum_{i=1,2}\frac{\sigma_i\,\mu_i B}{1+\mu_i^2B^2}.
    \label{eqn:sigma2band}
\end{equation}
Inverting this tensor, the Hall resistance takes the form
\begin{equation}
    R_{xy}(B)=B\,\frac{R_H(0)+R_H(\infty)\left(B/B^*\right)^2}{1+\left(B/B^*\right)^2},
    \label{eqn:Rxy2band}
\end{equation}
where $R_H(0)$ and $R_H(\infty)$ are the low- and high-field limits of the Hall coefficient $R_H(B)=R_{xy}/B$, and the crossover field $B^*$ is of order the inverse mobility of the more mobile carrier. Hall data therefore determine only three combinations of the four carrier parameters, and we fix the fourth with the measured zero-field sheet conductance, $\sigma_1+\sigma_2=1/R_\square$. Fits of Eq.~(\ref{eqn:Rxy2band}) give $B^*\approx0.6$--1.9 T. With $R_\square$ fixed, this curvature is reproduced only by a minority channel that carries a fraction $f\sim10^{-5}$ of the conductance with a mobility of 0.5--1.7 m$^2$/Vs. For $f\gtrsim10^{-4}$, the best two-carrier fit is no better than a straight line.

Mobilities of this size are characteristic of bulk FeSb$_2$ carriers \cite{Takahashi2011}, so bulk conduction is the natural candidate for the minority channel. The key test is the temperature dependence. The bulk of FeSb$_2$ is extraordinarily robustly insulating: its conduction remains thermally activated, with no sign of impurity-band or other parallel bulk conduction, down to the temperatures at which surface conduction takes over \cite{Eo2023}. Extrapolating this activated behavior to lower temperatures is therefore not an assumption of convenience, and it is consistent with the bulk line in Fig.~\ref{fig:four}a. A bulk minority channel must then follow
\begin{equation}
    \sigma_1(T)=\sigma_1^0\,\exp\!\left(-\frac{T_0}{T}\right).
    \label{eqn:activated}
\end{equation}
The most appropriate comparison is the low-temperature activation of the bulk Hall coefficient of FeSb$_2$, $E_{low}=6.26$ meV, which corresponds to $T_0\approx73$ K \cite{Eo2023}. For this value, the bulk conductance falls by a factor of $\approx4\times10^{8}$ between 3.5 and 1.8 K. Even the smaller slope of the bulk line in Fig.~\ref{fig:four}a ($T_0\approx50$ K, a lower bound because the surface channel already shunts the measured resistance in that range) gives a factor of $\approx7\times10^{5}$. By contrast, the minority conductance required by the free fits is constant to within a factor of $\approx2.5$ (Fig.~\ref{fig:S6}b), and the nonlinearity is largest at the lowest temperature. A global fit of all four temperatures that imposes Eq.~(\ref{eqn:activated}), with a shared minority mobility, cannot reproduce the curvature.

For comparison, an ordinary Hall term plus a saturating component, $R_{xy}=R_0B+A\tanh(B/B_s)$, describes the data equally well ($\chi^2\approx1.06$) with the same number of parameters, using only the measured low-mobility surface channel. Its saturation field is $B_s\approx0.2$--1 T, and its amplitude $A$ grows on cooling (Fig.~\ref{fig:S6}c).
}

\bibliography{references.bib}

\end{document}